\documentclass[reprint,aps,showpacs,floatfix,citeautoscript,longbibliography]{revtex4-1}

\pdfoutput=1

\usepackage{amsmath, amsfonts, amssymb, amsbsy}
\usepackage{graphicx}

\usepackage[caption=false]{subfig}

\makeatletter
\def\tagform@#1{\maketag@@@{\ignorespaces#1\unskip\@@italiccorr}}
\makeatother

\usepackage{hyperref}
\hypersetup{
        pdftitle={Comment on ''Probing two- and three-dimensional electrons in MgB2 with soft x-ray angle-resolved photoemission''},
        pdfdisplaydoctitle=true,
        pdfsubject = {Comment},
        pdfauthor= {BMW},
        pdfkeywords= {},
        colorlinks=true,
        linkcolor=blue,
        citecolor=blue,
        urlcolor=blue,
        unicode=false
}

\newlength{\circlediam}
\newcommand{\setcirclediam}{\setlength{\circlediam}{1.08\fontdimen6\font}}
\usepackage{tikz}
\newcommand*\circled[1]{\setcirclediam\tikz[baseline=(char.base)]{\node[shape=circle,draw,inner sep=0pt,minimum size=\the\circlediam] (char) {\textsc{#1}};}}

\usepackage{color}
\newcommand{\red}[1]{{\color{red}{#1}}}
\newcommand{\blue}[1]{{\color{blue}{#1}}}

\begin{document}
\title{Comment on ``Probing two- and three-dimensional electrons in MgB$_2$ with soft x-ray angle-resolved photoemission''}

\author{B.~M. Wojek}
\homepage{http://bastian.wojek.de/}
\affiliation{KTH Royal Institute of Technology, ICT MNF Materials Physics, Electrum 229, 164 40 Kista, Sweden}

\date{2 December 2015}

\begin{abstract}
A recent article by Sassa \emph{et al.} [\href{http://dx.doi.org/10.1103/PhysRevB.91.045114}{Phys. Rev. B \textbf{91}, 045114 (2015)}] reports on a 
soft x-ray angle-resolved photoemission study of MgB$_2$. The analysis and/or presentation of the collected data and the corresponding calculations
appear to be partially inconsistent. The aim of this comment is to provide a guide to these inconsistencies and to discuss their influence on the presented conclusions.
\end{abstract}

\pacs{71.20.-b, 79.60.-i}

\maketitle

\section{Introduction}

In a recent article Sassa \emph{et al.} presented a soft-x-ray angle-resolved photoemission (SX-ARPES) study of MgB$_2$~\cite{Sassa-PhysRevB-2015},
claiming to have established ``the full 3D electronic structure'' of the compound. The experimental data and the results of band-structure 
calculations are compared and certain deviations, e.g., for the band width and Fermi-surface area are reported. However, the illustrations used to circumstantiate the article's quantitative claims seem to be mutually inconsistent. In the following, firstly these inconsistencies will be described in detail (but not necessarily in any hierarchical order). Secondly, their influence on the conclusions of Ref.~\onlinecite{Sassa-PhysRevB-2015} is discussed.

\section{Observations}
\label{sec:observations}
The main figure (Fig.~5~\cite{figures}) involved in the band-structure analysis contains a set of particular oddities. All the \emph{calculated} bands 
depicted in Figs.~5(a) and~5(c) ought to be symmetric about $M$ and $L$---evidently, some are not: in Fig.~5(a) the $\pi$ band from $\Gamma$ to $M$ is 
farther away from $M$ than the $\pi$ band beyond $M$, leading to differences in the crossing points with the $\sigma$ bands ($1.67$~eV vs. $1.58$~eV 
and $2.47$~eV vs. $2.37$~eV). In Fig.~5(c) the second $\pi$ band and the third $\sigma$ band (always counting at $L$ from low to high binding energy 
$E_\mathrm{B}$) are shifted significantly toward $A$. The top of both bands is found at about $-1.13~\pi/a$ instead of 
$-2/\sqrt{3}~\pi/a$~\cite{labels}. Hence, also the crossings of the second $\pi$ and second $\sigma$ bands are not at a constant $E_\mathrm{B}$ 
($5.01$~eV vs. $4.86$~eV). Apparently, this is \emph{not} an effect of inaccuracies of the calculations, otherwise, the second band shown in 
the zoom-in Fig.~5(g) would not be centered at $L$, either.

\begin{figure}
\centering
\includegraphics[width=\columnwidth]{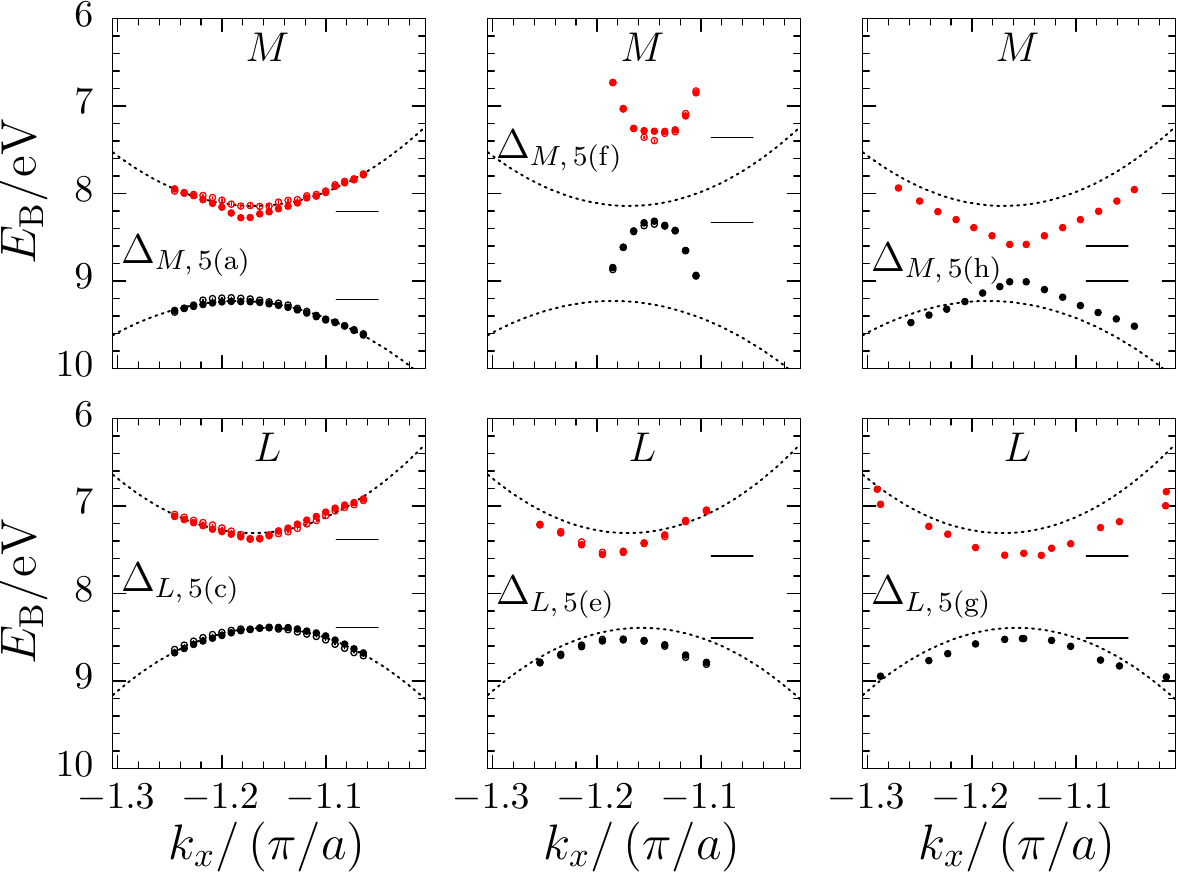}
\caption{\emph{Upper panels}: Peak positions from two-peak EDC fits to the data around the $M$ point in Figs.~5(a) and~5(f) and the corresponding 
positions depicted in Fig.~5(h). \emph{Lower panels}: Peak positions from two-peak EDC fits to the data around the $L$ point in Figs.~5(c) and~5(e) 
and the corresponding positions depicted in Fig.~5(g). Filled and open symbols are for fits using coupled and individual Lorentzian 
widths for the EDCs. The central panels assume an equidistant spacing of the EDCs based on the labels given in the Figs.~5(f) and~5(e). The dotted curves represent parabolic fits to the peak positions in the left-most panels at $M$ and $L$, respectively. They are reproduced in the middle and right panels to facilitate a better comparison between the different presentations.}
\label{fig:EDCpeaks}
\end{figure}

Staying with Fig.~5 and reviewing the energy-distribution-curve (EDC) analysis reveals several inconsistencies between the presentations in the 
various subfigures. According to the text in Ref.~\onlinecite{Sassa-PhysRevB-2015}, the EDCs are analyzed using ``resolution broadened Lorentzian 
functions''. Following the same route, here, EDCs around $M$ obtained from the grayscale map in Fig.~5(a) and the (assumed equally spaced) EDCs in 
Fig.~5(f) are modeled with a pair of Voigtians [Gaussian-broadened ($\sigma=0.1$~eV) Lorentzians] residing on a polynomial background [either 
parametrized from Fig.~5(f) or fitted]. Two fits are performed for each set of data: using a $E_\mathrm{B}$ fitting range between $5.5$~eV and 
$10.5$~eV firstly the curves are modeled with two Lorentzian widths that are shared by all EDCs and secondly the fit is repeated with two independent 
Lorentzian widths for each EDC. Both procedures yield fits that are comparable to those shown in Fig.~5(f). In the first case the obtained Lorentzian 
half-widths at half maximum (HWHM) are $0.6$~eV and $0.9$~eV, respectively. In the second case, the HWHM range from $0.1$~eV (with small 
corresponding amplitudes) to $0.6$~eV and from $1.1$~eV to $1.2$~eV, respectively. The obtained EDC peak positions are shown in 
Fig.~\ref{fig:EDCpeaks} and compared to the peak positions marked in Fig.~5(h). If the presentation was consistent throughout Fig.~5, the curves 
sampled by the peak positions shown in the three upper panels of Fig.~\ref{fig:EDCpeaks} would coincide. What is observed instead is that the 
presentation in Fig.~5(f) is shifted in energy by about $0.85$~eV with respect to the data in Figs.~5(a) as well as~5(h) and that it features 
erroneous labeling; the top-most EDC in Fig.~5(f) appears to be rather at $k_x\approx-1.00~\pi/a$ or $k_x\approx-1.30~\pi/a$ [if the same decreasing 
$k_x$---not $k_y$---as in Fig.~5(e) is used for higher up EDCs]. In addition, while the band gap for the analysis of the data presented in Figs.~5(a) 
and~5(f) is consistently sized about $0.9$~eV to $1.0$~eV, the purported peak positions shown in Fig.~5(h) yield a gap of $0.4$~eV only. Moreover, 
both gap labels in Fig.~5(h) seem to be inadequate: to be consistent with the points shown in this subfigure as well as the number given in the main 
text (ignoring for the moment the variance with the other subfigures) the ``experimental gap'' should equal $0.4$~eV; the gap between the calculated 
curves is $0.6$~eV also at the $M$ point. However, taking into account all available information, the experimental gap at $M$ seems more likely to 
have a size of about $0.9$~eV to $1.0$~eV---thus being larger than calculated and at major variance with the peak positions depicted in Fig.~5(h).

An analogous analysis for the data close to $L$ presented in Figs.~5(c), 5(e), and~5(g) yields an overall much more consistent picture as shown in 
the lower panels of Fig.~\ref{fig:EDCpeaks}. Also here, a comparably small shift in energy of about $0.2$~eV might be present between the data in 
Fig.~5(c) and the other two subfigures, yet, the overall dispersions and gap sizes match rather well as they should.

\begin{figure}
	\centering
	\includegraphics[width=.65\columnwidth]{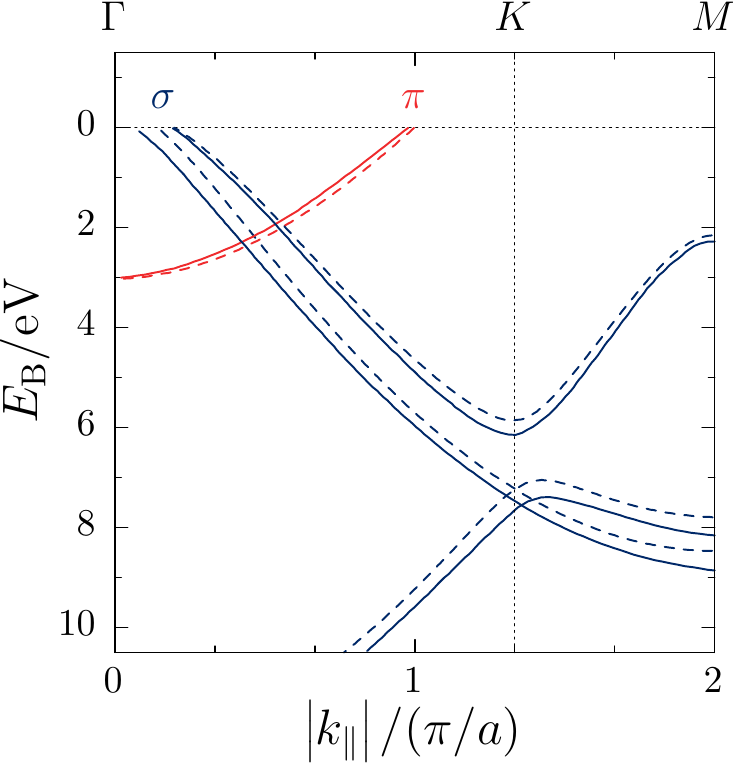}
	\caption{Comparison of the calculated band structures depicted in Figs.~4 (dashed lines) and~5(b) (solid lines).}
	\label{fig:compareGKMplots}
\end{figure}

The calculated band structure along the $\Gamma$-$K$-$M$ high-symmetry line is shown twice in Ref.~\onlinecite{Sassa-PhysRevB-2015}: once in Fig.~5(b) and as well in the left panel of Fig.~4. Naturally, both presentations are expected to show identical results. However, when plotted together in Fig.~\ref{fig:compareGKMplots}, it is evident that there are nonnegligible differences between the two presentations. The deviations in the depicted band positions partially reach up to $0.4$~eV or more than $0.05\,\pi/a$.

The data in Fig.~5 is also used to compare variations in the band width between the experiment and the calculation. For instance, ``(f)or the 
$\sigma$ bands, the calculation gives a width that is $\sim{}8\,\%$ smaller in the $A$-$L$ than in the $\Gamma$-$M$ direction. The experiment, 
however, shows the same bandwidth for both of these direction(s) (...)''~\cite{Sassa-PhysRevB-2015}. Again, using the data from Figs.~5(a) and~5(c) it 
is possible to estimate the bottom of the first two $\sigma$ bands and thus their band width. The bottom of the first $\sigma$ band is at $2.43$~eV at $M$ 
and at $2.06$~eV at $L$, the bottom of the second $\sigma$ band is at $8.21$~eV at $M$ and at $7.38$~eV at $L$. Thus, the \emph{experimental} $\sigma$ band width is reduced 
by $10\,\%$ to $15\,\%$ for $A$-$L$ compared to $\Gamma$-$M$. Furthermore, using the results of the \emph{calculations} shown in the same figures, one finds the bottom of the first $\sigma$ band at $2.31$~eV at $M$ 
and at $1.91$~eV at $L$. The bottom of the second $\sigma$ band is at $8.10$~eV at $M$ and at $6.95$~eV at $L$. Hence, the presented calculation actually suggests a reduction of the band width by $14\,\%$ to $17\,\%$ for $A$-$L$ compared to $\Gamma$-$M$. Therefore, the experiment seems to be in full qualitative agreement with the shown calculations. Yet, neither the supposed observed equal experimental band width nor the expected $\sim{}8\,\%$ band-width reduction quoted above can be deduced from the figures.

Another contestable practice used in Figs.~5 and~3 is the disregard of $k_z$ variations for constant photon energies when $k_\parallel$ becomes 
comparably large, despite quoting Eq.~(1) in Ref.~\onlinecite{Sassa-PhysRevB-2015}. For instance, in the used photon-energy range in the vicinity of 
the Fermi energy $k_z$ changes by almost $0.1\,\pi/c$ when increasing $k_\parallel$ from $0$ to $2/\sqrt{3}\,\pi/a$. This implies that some 
caution is in order when comparing constant-photon-energy measurements with constant-$k_z$ calculations, even more so if bands with strong $k_z$ 
dispersion like the $\pi/\pi^{\ast}$ bands of MgB$_2$ are studied. Despite the claim of Ref.~\onlinecite{Sassa-PhysRevB-2015} to have revealed the 
``complete 2D and 3D electronic band structure of MgB$_2$'', with the exception of parts of the Fermi surface and bands at two distinct $k_z$ points, 
\emph{no} out-of-plane dispersion has been presented. Therefore, it is difficult to assess quantitatively the effects of the $k_z$ variations on the 
measured band structure within the calculations of Ref.~\onlinecite{Sassa-PhysRevB-2015}. However, as evident from the calculated Fermi surface, expected
qualitative effects include, e.g., the diminishing of the $L$ electron band close to the Fermi energy in Fig.~5(c) and the corresponding growth of the
``$\pi$ Fermi hexagons'' in Fig.~3(b) when $k_z$ is moving away from $(11)\,\pi/c$.

\begin{table*}
	\caption{\label{tab:pos_FS} Comparison of the absolute values of the \emph{calculated} in-plane components of the Fermi crystal momenta in units of $\pi/a$ along the different high-symmetry lines shown in Figs.~1, 2(d), 2(e), 3, 4, and 5 of Ref.~\onlinecite{Sassa-PhysRevB-2015}. Using the plots in Fig.~3 as reference, values deviating by more than $0.02\,\pi/a$ ($0.05\,\pi/a$) are marked in \blue{blue} (\red{red}).}
	\begin{ruledtabular}
		\begin{tabular}{ccccccccc}
			& $k_{\mathrm{F},1}^{\parallel}$ & $k_{\mathrm{F},2}^{\parallel}$ & $k_{\mathrm{F},3}^{\parallel}$ & $k_{\mathrm{F},4}^{\parallel}$ & $k_{\mathrm{F},5}^{\parallel}$ & $k_{\mathrm{F},6}^{\parallel}$ & Figure & Notes\tabularnewline
			\noalign{\smallskip}
			\hline
			\noalign{\smallskip}
			$\Gamma$-$M$-$\Gamma$ & $2.16$ & $2.10$ & $1.35$ & $0.96$ & $0.21$ & $0.15$ & 3(c) & $k_y = 0$, $k_x < 0$ \tabularnewline
			                      & $2.16$ & \blue{$2.07$} & \red{$1.29$} & \red{$1.03$} & \blue{$0.24$} & $0.15$ & 1 & base plane, left \tabularnewline
			                      & \blue{$2.13$} & $2.09$ & \red{$1.22$} & \red{$1.04$} & $0.22$ & $0.17$ & 5(a) & \tabularnewline
			                      & --- & --- & \blue{$1.32$} & \blue{$0.99$} & \red{$0.31$} & \blue{$0.19$} & 2(d) & $k_z=16\,\pi/c$, $k_x < 0$ \tabularnewline
   			\noalign{\smallskip}
   			\hline
   			\noalign{\smallskip}
   			$M$-$K$-$\Gamma$ & --- & --- & --- & $0.89$ & $0.20$ & $0.15$ & 3(c) & $k_x = 0$, $k_y < 0$ \tabularnewline
   			                 & --- & --- & --- & \red{$0.95$} & $0.22$ & $0.15$ & 1 & top plane, back $\Gamma$ \tabularnewline
   			                 & --- & --- & --- & \red{$0.98$} & $0.19$ & \red{$0.07$} & 5(b) & \tabularnewline
   			                 & --- & --- & --- & \red{$0.98$} & --- & --- & 2(e) & $k_z=16\,\pi/c$, $k_y < 2\,\pi/a$\tabularnewline
   			                 & --- & --- & --- & \red{$1.00$} & $0.20$ & $0.14$ & 4 & \tabularnewline
   			\noalign{\smallskip}
   			\hline
   			\noalign{\smallskip}
   			$A$-$L$-$A$ & $2.09$ & $1.99$ & $1.43$ & $0.89$ & $0.31$ & $0.22$ & 3(d) & $k_y = 0$, $k_x < 0$ \tabularnewline
   			            & $2.08$ & $1.97$ & \blue{$1.39$} & \blue{$0.92$} & \blue{$0.34$} & $0.24$ & 1 & central plane, left \tabularnewline
   			            & \blue{$2.13$} & \blue{$2.02$} & $1.41$ & $0.89$ & \blue{$0.28$} & \blue{$0.18$} & 5(c) & \tabularnewline
   			            & --- & --- & \red{$1.30$} & \red{$1.01$} & \red{$0.44$} & \blue{$0.27$} & 2(d) & $k_z=15\,\pi/c$, $k_x < 0$\tabularnewline
   			\noalign{\smallskip}
   			\hline
   			\noalign{\smallskip}
   			$L$-$H$-$A$ & --- & --- & --- & $0.93$ & $0.30$ & $0.22$ & 3(d) & $k_x = 0$, $k_y < 0$ \tabularnewline
   			            & --- & --- & --- & \red{$1.01$} & $0.32$ & $0.24$ & 1 & central plane, left \tabularnewline
             		    & --- & --- & --- & \red{$0.99$} & $0.31$ & \blue{$0.25$} & 5(d) & \tabularnewline
   		                & --- & --- & --- & \red{$0.99$} & --- & --- & 2(e) & $k_z=15\,\pi/c$, $k_y < 2\,\pi/a$\tabularnewline
		\end{tabular}
	\end{ruledtabular}
\end{table*}

Focusing now on the $k_z$-dependent cuts through the Fermi surface shown in Figs.~2(a) through~2(e), also here, inconsistencies in the presentation are found. 
Figures~2(d) and~2(e) feature common $k_z$ tic marks and both cuts \protect\circled{1} and \protect\circled{2} include the $M$-$L$ line. Hence, the intersections 
of this line with the \emph{calculated} Fermi surface parts should be identical in Figs.~2(d) and~2(e). Surprisingly, they are not. Going from high $k_z$ to 
low $k_z$, the corresponding $k_z/(\pi/c)$ values are $16.56$ vs. $16.64$, $16.25$ vs. $16.29$, $15.75$ vs. $15.65$, $15.42$ vs. $15.29$, $14.58$ vs. 
$14.61$, $14.25$ vs. $14.25$, $13.75$ vs. $13.61$, and $13.43$ vs. $13.26$. Moreover, the shown curves are overall not symmetric about the $k_z$ 
integer multiples of $\pi/c$. Maybe most prominently, this is seen at the orange curve off-centered from $k_z=14\,\pi/c$ in Fig.~2(e) as well as the
cross section of the warped $\sigma$ Fermi-surface cylinders in Fig.~2(d) which are not ``narrowest at $\Gamma$'' as stated in the text of Ref.~\onlinecite{Sassa-PhysRevB-2015}.

Since the presentations of the calculated bands in Fig.~5 as well as the Fermi surface in Figs.~2(d) and~2(e) contain apparent inconsistencies as detailed above, the overall
positions of the in-plane components of the \emph{calculated} Fermi crystal momenta along different high-symmetry lines throughout the various illustrations in
Ref.~\onlinecite{Sassa-PhysRevB-2015} are compared in Table~\ref{tab:pos_FS}. Deviations of partially more than $0.1\,\pi/a$ can be observed between the different figures.

Further inconsistencies can be found in Figs.~2(a) through~2(e). Illustrated in Fig.~2(a) and consistent with the labels in Figs.~2(d) and~2(e), cuts \protect\circled{1} and
\protect\circled{2} lie in the $k_x$-$k_z$ and $k_y$-$k_z$ planes, respectively. Contrarily, it is stated in the figure caption that both cuts cover the 
$k_x$-$k_z$ plane. Moreover, the introduction of Ref.~\onlinecite{Sassa-PhysRevB-2015} advertises that for reporting the experimental results ``data 
enhancement (was) not needed''. This is at variance with the apparent but not indicated symmetrized presentation of experimental data in Figs.~2(b) 
and~2(c). It might also be worth noting that the part of the Fermi surface around $L$ which is deemed ``not visible'' is indeed present in the 
data---only due to an erroneous symmetrization of the data vanishing in cut \protect\circled{2}, but nevertheless constituting the strongest signal 
in cut \protect\circled{1}. A detailed account of the effects of the apparent data treatment in Figs.~2 and~3 can be found in the 
appendix.

Finally, a few minor issues in the experimental description of Ref.~\onlinecite{Sassa-PhysRevB-2015} are identified. It has been mentioned that ``by 
scanning the photon energy ($h\nu$) at fixed binding energy $E_\mathrm{B}$ and emission angle'' one could scan, e.g., along $k_\parallel=0$. Due to 
the finite momentum transfer from the incident photons this is not strictly the case: $k_\parallel=0$ does not correspond to electrons emitted normally to the sample surface
and the emission angle changes continuously with the photon energy/momentum. Also, the correction term taking into account the photon 
momentum transfer for the calculation of $k_z$ in Eq.~(1) lacks a division by $\hbar$. Eventually, while the angle between the analyzer axis and the 
incident photon beam is $70^\circ$ at the ADRESS SX-ARPES end station, the angle used in Eq.~(1) ought to be the one between the inclined 
analyzer axis and the laboratory normal ($20^\circ$)~\cite{Strocov-calculations}.

\section{Discussion}
Having identified a multitude of inconsistencies in the presentation of the results, an important question is how these impact the 
analysis and conclusions of Ref.~\onlinecite{Sassa-PhysRevB-2015}. This question should be addressed in two parts: Firstly, 
Ref.~\onlinecite{Sassa-PhysRevB-2015} establishes experimentally the topology of the 3D Fermi-surface as well as of the \emph{in-plane} dispersion of 
MgB$_2$. The observed features are qualitatively well reproduced by calculations in the literature (cf., e.g., references within 
Ref.~\onlinecite{Sassa-PhysRevB-2015}). While some of the presentations of experimental data are indeed questionable (see appendix), the apparent 
flaws described above do not affect this qualitative result. The second part of the question concerns if the detailed quantitative comparisons between 
the ARPES data and the presented calculations still hold despite the inconsistencies within and across the different illustrations. To answer that a 
few concrete examples shall be discussed in the following.

Ref.~\onlinecite{Sassa-PhysRevB-2015} has a particular emphasis on deviations of the $\pi/\pi^{\ast}$ calculations from the measured bands. For 
instance, ``(a)t $L$, the measured electron pocket has a smaller FS cross section (by $\sim 10\%$) than what is estimated by (the) calculation [Fig. 
5(c)]. (\dots) To correct for this, the calculated antibonding $\pi^{\ast}$ band would have to be shifted by approximately $200\pm50$~meV toward 
$E_{F}$.'' This is at substantial variance with the Fermi surface shown in Fig.~2(d) (cf. $k_{\mathrm{F},3}^{\parallel}$ and 
$k_{\mathrm{F},4}^{\parallel}$ in Table~\ref{tab:pos_FS}). To reproduce the values of $k_{\mathrm{F},3}^{\parallel}$ and 
$k_{\mathrm{F},4}^{\parallel}$ in Fig.~2(d), this very band would have to be shifted by $0.8$~eV toward $E_{\mathrm{F}}$---multiple times more than 
the deviation discussed here. Although in fact this correction seems unlikely and would probably be at variance with the $k_z$ dispersion of the band, 
it nonetheless illustrates the order of magnitude of the observed inconsistencies in Ref.~\onlinecite{Sassa-PhysRevB-2015} and thus a substantial 
uncertainty introduced by them.

Next, ``(f)or the $\pi$ bands, the measured width for the band dispersing from $L$ to $H$ to $A$ [Fig. 5(d)] is $\sim10$--$15$~\% larger than the 
calculated one''~\cite{Sassa-PhysRevB-2015}. Without taking into account the argument just made and assuming the illustration in Fig.~5(d) represents 
the valid result of the calculation for $k_z=\pi/c$, the consequences of the $k_z$ variation with changing $k_y$ in the data can be estimated. As 
mentioned above, Ref.~\onlinecite{Sassa-PhysRevB-2015} does not provide out-of-plane dispersions. Therefore, calculations of 
Ref.~\onlinecite{Singh-PhysRevLett-2001} for the $L$-$M$ direction are used for an estimate. According to Eq.~(1) $k_z$ varies by about $0.24~\pi/c$ 
between $L$ ($k_y=\pm2~\pi/a$) and $A$ ($k_y=0$) for $h\nu=370$~eV and energies close to the Fermi energy. Moving $0.24~\pi/c$ away from $L$ toward 
$M$, changes the topmost occupied energy level by about $0.4$~eV toward $E_\mathrm{F}$~\cite{Singh-PhysRevLett-2001} which corresponds to about 
$6.5$~\% of the band width. Hence, the deviation reported in Ref.~\onlinecite{Sassa-PhysRevB-2015} is potentially overestimated by about a factor of 
$2$.

Already in the previous section the comparison of the $\sigma$ band width between $A$-$L$ and $\Gamma$-$M$ as seen in Figs.~5(a) and~5(c) has been 
discussed in detail. It is emphasized once more that neither the supposed observed equal experimental band width along these directions nor the 
expected $\sim{}8\,\%$ band-width reduction for $A$-$L$ compared to $\Gamma$-$M$ quoted in Ref.~\onlinecite{Sassa-PhysRevB-2015} can be deduced from 
the figures.

The detailed analysis presented around Fig.~\ref{fig:EDCpeaks} suggests that contrary to the claim of Ref.~\onlinecite{Sassa-PhysRevB-2015} the band 
gap at $M$ at high binding energies is most likely not smaller than calculated, but rather has a similar size than the deduced band gap at $L$ and 
thus is somewhat larger than predicted. It is also worth noting here, that this band gap is expected to have a negligible $k_z$ 
dependence~\cite{Singh-PhysRevLett-2001}, which---in contrast to the findings of Ref.~\onlinecite{Sassa-PhysRevB-2015}---is nicely confirmed in 
Fig.~\ref{fig:EDCpeaks}.

In connection with Figs.~5(g) and~5(h) it is also pointed out in Ref.~\onlinecite{Sassa-PhysRevB-2015} that ``a shift of the calculated band(s) edges 
by approximately $0.6$ eV is needed in order to fit the data (\dots)''. Given the uncertainty of the calculations arising from the comparison of the 
calculated band structures depicted in Figs.~5(b) and 4, the relevant energy levels at $M$ might be shifted up to $0.4$~eV to lower binding energies, 
thus increasing the postulated deviation by $2/3$ to about $1.0$~eV (keeping the originally marked peak positions as reference).

Statements like ``the $\sigma$ band width `is $1$--$2$~\% wider than the calculated one along $\Gamma$-$M$'~\cite{Sassa-PhysRevB-2015}'' depend on a 
very high degree of accuracy in the calculations. The uncertainty originating from Fig.~\ref{fig:compareGKMplots}---which easily amounts up to $5$~\% 
of the band width---renders such conclusions not reliable.

Eventually, the host of significant variations in the calculated Fermi crystal momenta summarized in Table~\ref{tab:pos_FS} as well as the overall 
uncertainty in the presented band calculations discussed in the preceding examples most likely preclude a meaningful detailed quantitative comparison 
between the experimental data and the calculations. Therefore, also the comparison of the conclusions obtained in 
Ref.~\onlinecite{Sassa-PhysRevB-2015} with previous de Haas--van Alphen Fermi-surface measurements has to be seen overall critically.

Taking into account all the effects discussed above, it is concluded that the results of the quantitative analysis in 
Ref.~\onlinecite{Sassa-PhysRevB-2015} are indeed substantially influenced by the inconsistencies reported here. With the only exception of the 
band-gap determination at the $L$ point, the quantitative results are found to be questionable or at least subject to major uncertainty which could 
only be reduced by a fully consistent presentation.

\section{Summary}
In summary, the article by Sassa \emph{et al.} in principle provides experimental data of high quality. Some of the corresponding presentations are 
rather questionable (see appendix) but any qualitative arguments regarding the topology of the 3D Fermi surface and of the in-plane electronic band 
structure of MgB$_2$ appear to be valid. Yet, a multitude of inconsistencies in the analysis and/or presentation of the results cast doubt on the 
reliability of the obtained quantitative conclusions. Moreover, the fundamental claim of the article to have established ``the full 3D electronic 
structure'' of MgB$_2$ is found to be somewhat exaggerated.

\begin{acknowledgments}
I thank the staff of the ADRESS beamline at the Swiss Light Source (Paul Scherrer Institute, Villigen, Switzerland) for providing access 
to the experimental data discussed in the appendix of this comment.

The following free software was particularly useful while preparing this manuscript: Debian GNU/Linux~\cite{Debian} including KDE~\cite{KDE}, 
Inkscape~\cite{inkscape}, Engauge~\cite{engauge}, GNU Octave~\cite{octave}, ROOT~\cite{ROOT}, gnuplot~\cite{gnuplot}, and TeX Live~\cite{texlive} as 
well as musrfit~\cite{SuterWojek-PhysicsProcedia-2012}.

This work was supported by the Swedish Research Council.
\end{acknowledgments}

\appendix*
\section{Presentation of Fermi-surface data}
\label{sec:appendix}
It has already been mentioned in Sec.~\ref{sec:observations} that the Fermi-surface data shown in Ref.~\onlinecite{Sassa-PhysRevB-2015} are seemingly 
symmetrized. To shine some light on this procedure, in the following the illustrations in Figs.~2(b) and~2(c) as well as in Figs.~3(a) and~3(b) shall 
be compared to nonsymmetrized presentations derived from the experimental data~\cite{PSI}. The data analysis performed in this comment merely 
involves rudimentary data alignment and normalization steps as well as an \emph{approximate} mapping to reciprocal space. Slight further angular 
adjustments would be required for an exact quantitative analysis. However, this is beyond the scope of the qualitative comparison in this appendix.

\begin{figure}
\centering
\subfloat{\includegraphics[width=.485\columnwidth]{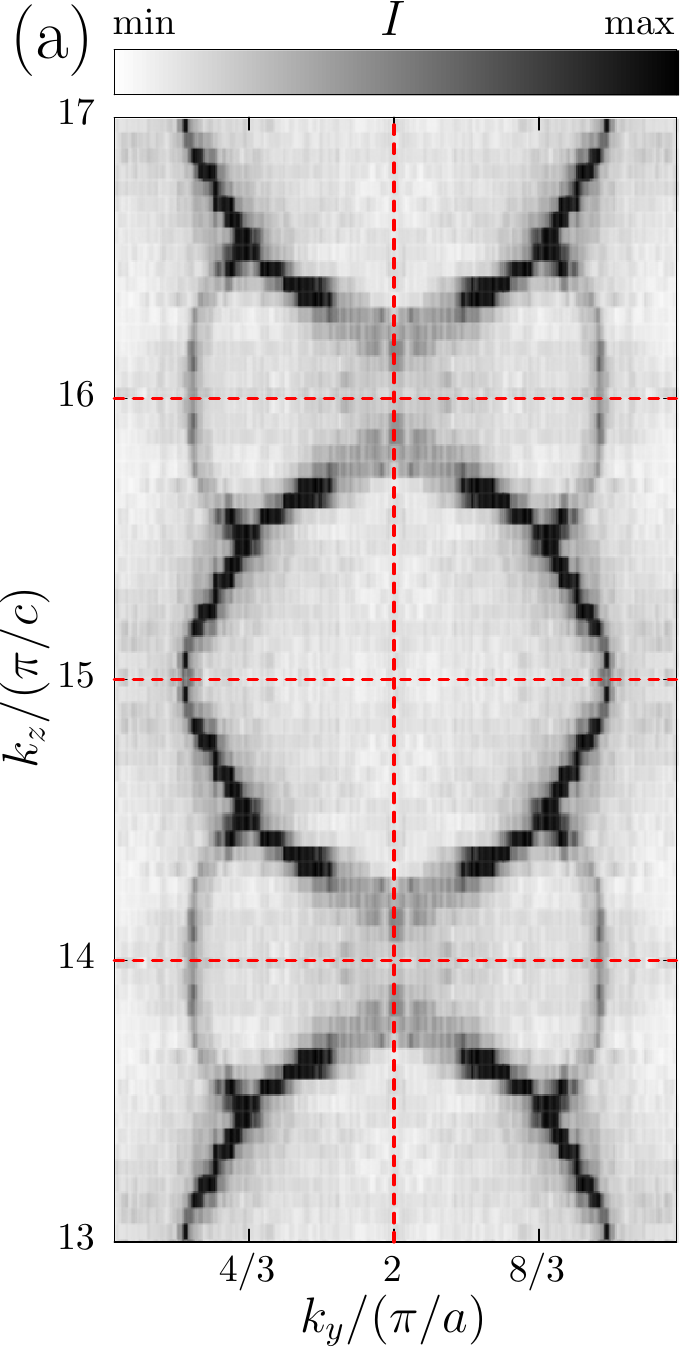}\label{fig:KyKz_GK:a}}\hspace*{.02\columnwidth}
\subfloat{\includegraphics[width=.485\columnwidth]{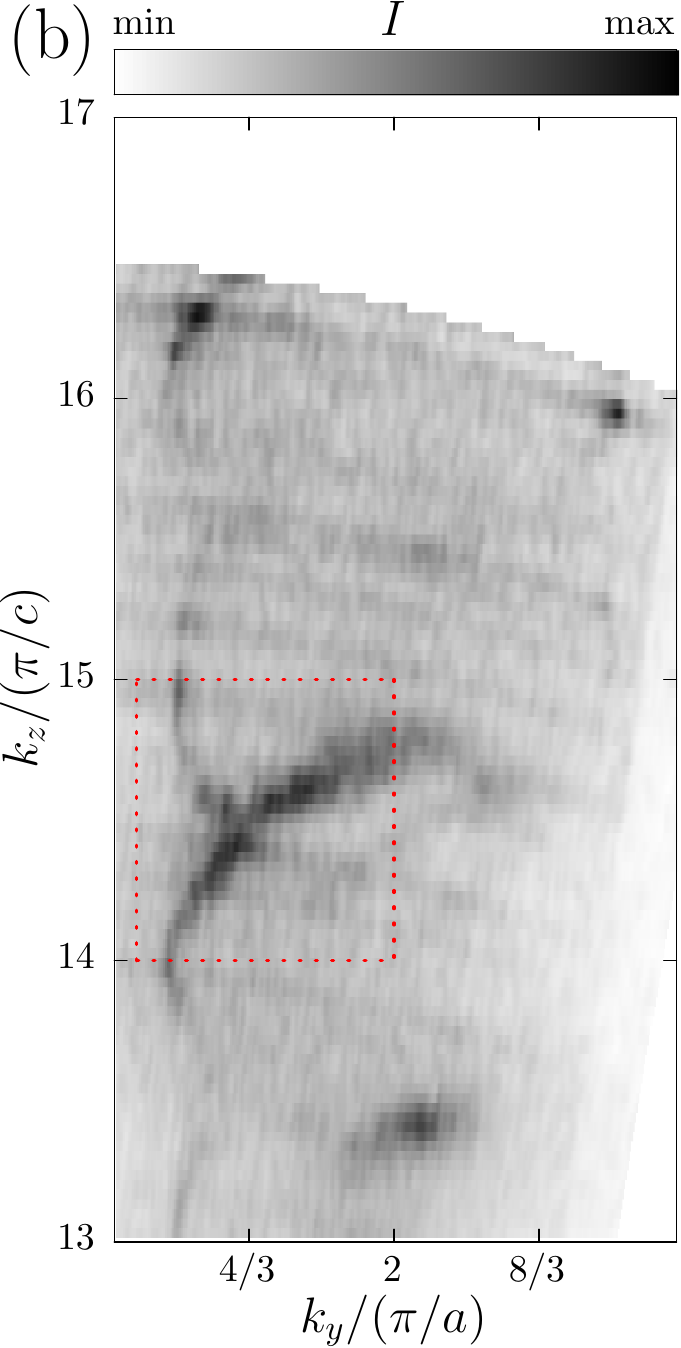}\label{fig:KyKz_GK:b}}
\caption{(a) Replotted presentation of Fig.~2(c) without decoration. The red dashed lines border the symmetrized sections of this illustration---only 
a single section appears to be the basis for the entire presentation. (b) Corresponding nonsymmetrized data acquired using circularly polarized 
light. The dotted rectangle approximately marks the part of the data used most probably for the symmetrization in (a).}
\label{fig:KyKz_GK}
\end{figure}

First, the experimental Fermi surfaces shown in Fig.~2 are analyzed. Figure~\ref{fig:KyKz_GK} compares the (redrawn) presentation of Fig.~2(c) 
[Fig.~\ref{fig:KyKz_GK:a}] with a direct mapping of the corresponding experimental data not involving any symmetrization, background subtraction, or 
the like [Fig.~\ref{fig:KyKz_GK:b}]. It seems to be clear that only a small part of the experimental data has been used as basis of the 
illustration in Fig.~2(c) [dotted rectangle in Fig.~\ref{fig:KyKz_GK:b}]. Moreover, it is apparent that the symmetrization in Fig.~2(c) is flawed: 
The data is shifted by $\pm{}\pi/c$ in $k_z$ direction, leading to the erroneous conclusion in Ref.~\onlinecite{Sassa-PhysRevB-2015} that the 
electron-like ``pocket centered around $L$ is not visible [red dashed line in Fig. 2(c)]''~\cite{Sassa-PhysRevB-2015}. Rather, it is a part of the 
hole-like pocket around $M$ which can hardly be observed in this experiment. This resolves the inconsistency with Fig.~2(b), where indeed the 
electron-like Fermi-surface around $L$ contributes the most intense photoelectron signal. The latter is also evident from a presentation of 
ARPES data acquired using $p$-polarized light in Fig.~1.1(b) of Ref.~\onlinecite{ETH-2011} (albeit with apparently questionable labeling on the 
$k_x$ axis).

\begin{figure}
\centering
\subfloat{\includegraphics[width=.485\columnwidth]{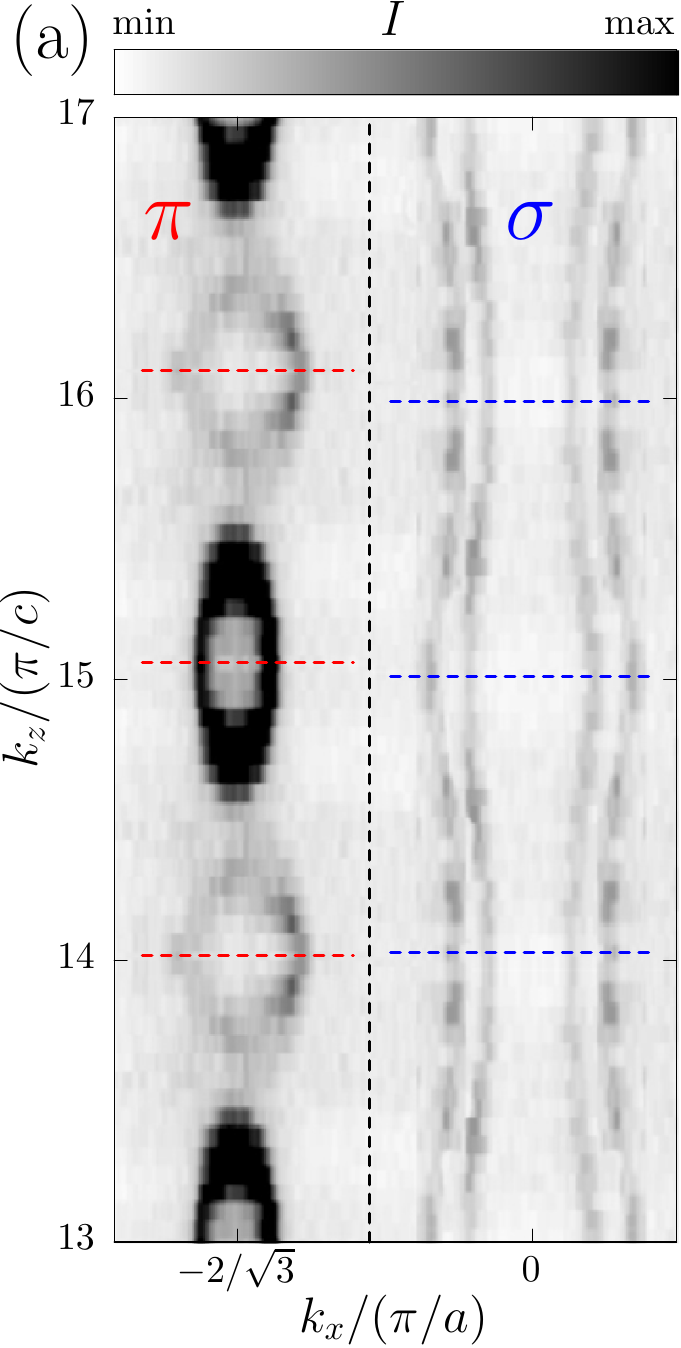}\label{fig:KxKz_GM:a}}\hspace*{.02\columnwidth}
\subfloat{\includegraphics[width=.485\columnwidth]{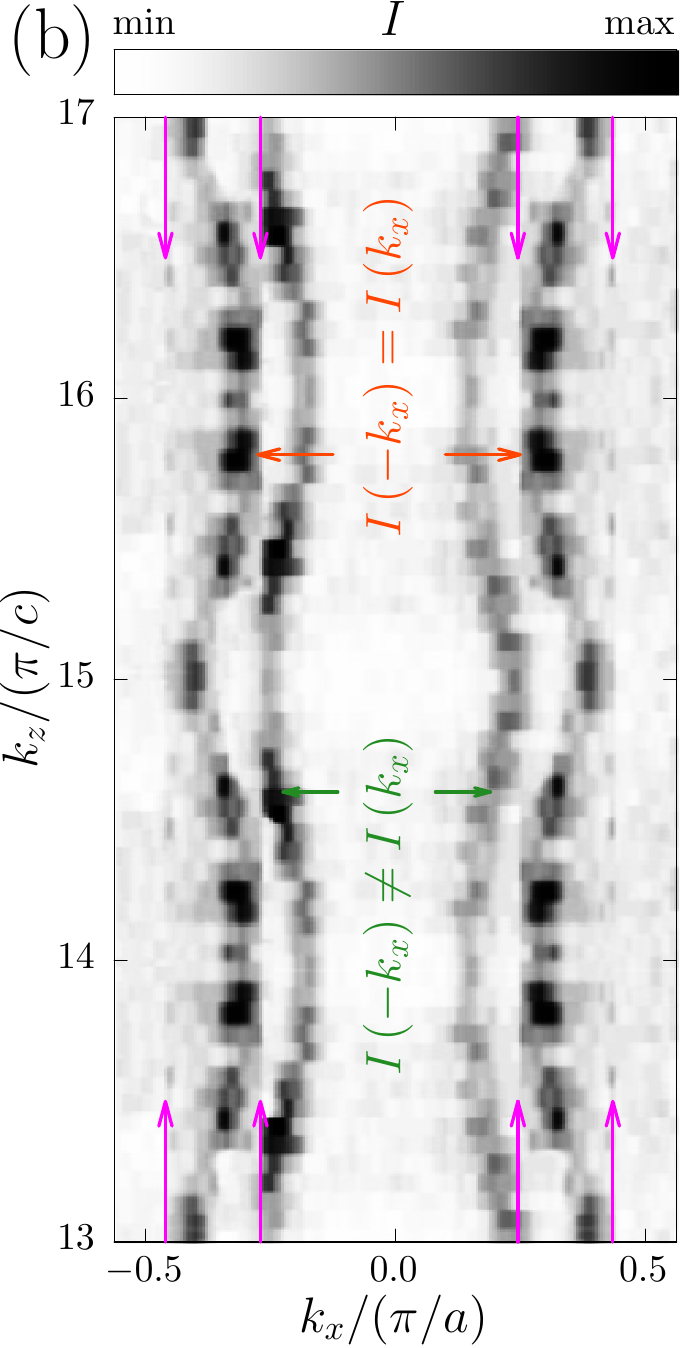}\label{fig:KxKz_GM:b}}\\
\subfloat{\includegraphics[width=.485\columnwidth]{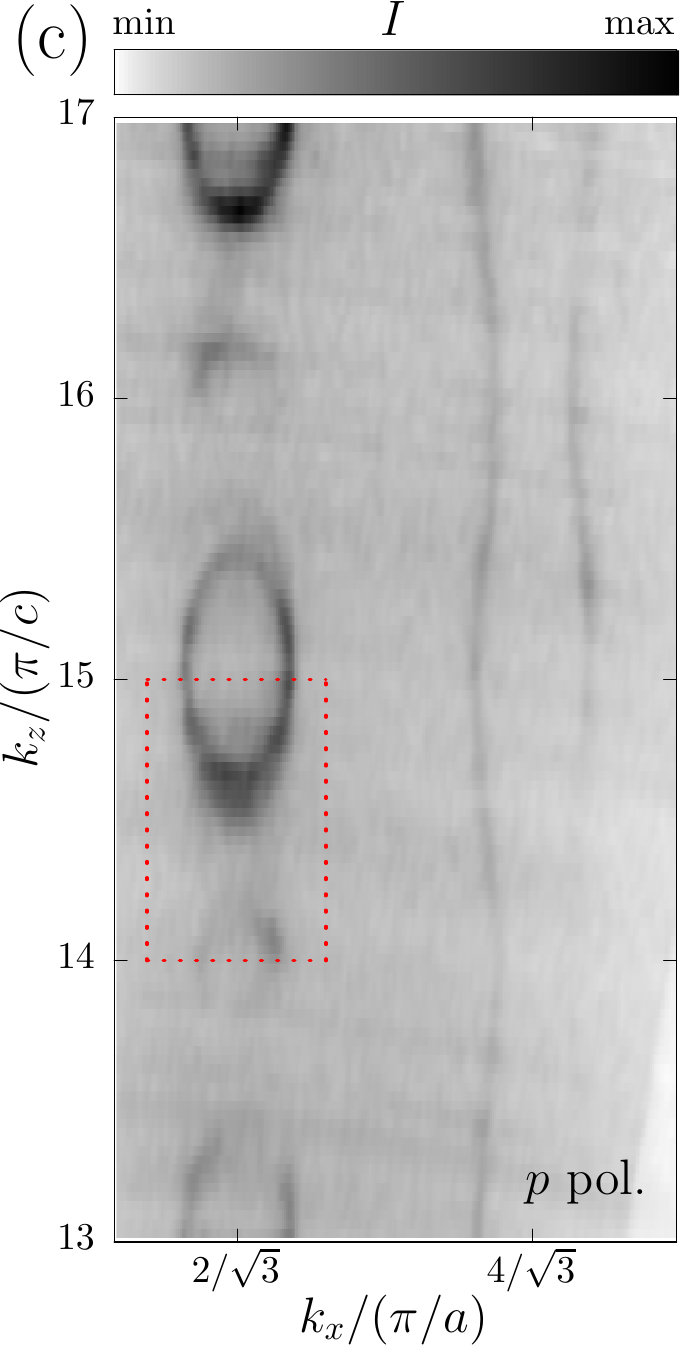}\label{fig:KxKz_GM:c}}\hspace*{.02\columnwidth}
\subfloat{\includegraphics[width=.485\columnwidth]{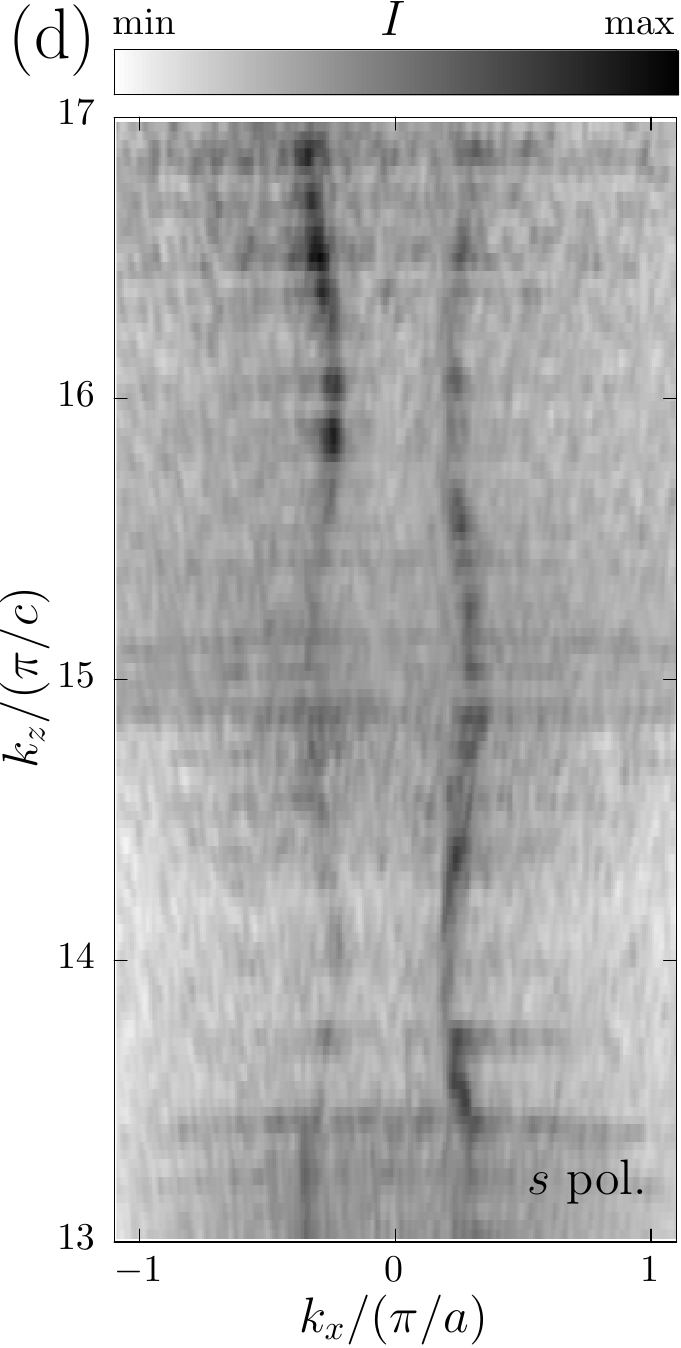}\label{fig:KxKz_GM:d}}
\caption{(a) Replotted presentation of Fig.~2(b) without decoration. The red and blue dashed lines indicate the symmetrized sections (half periods 
in $k_z$ direction) of this illustration. Apparently, the $\pi$ and $\sigma$ Fermi surfaces feature different periodicities. The black dashed line 
approximately marks the border between the regions with different $k_z$ periodicities. (b) Replotted $\sigma$ Fermi surfaces in 
a different grayscale than in (a). The purple arrows mark the visible discontinuities at $k_x = \text{const.}$ (c) Nonsymmetrized Fermi-surface data 
acquired using $p$-polarized light. The dotted rectangle approximately marks the part of the data reminiscent of the symmetrized $\pi$ data in 
(a). (d) Nonsymmetrized data acquired using $s$-polarized light.}
\label{fig:KxKz_GM}
\end{figure}

\begin{figure}
\centering
\subfloat{\includegraphics[width=.485\columnwidth]{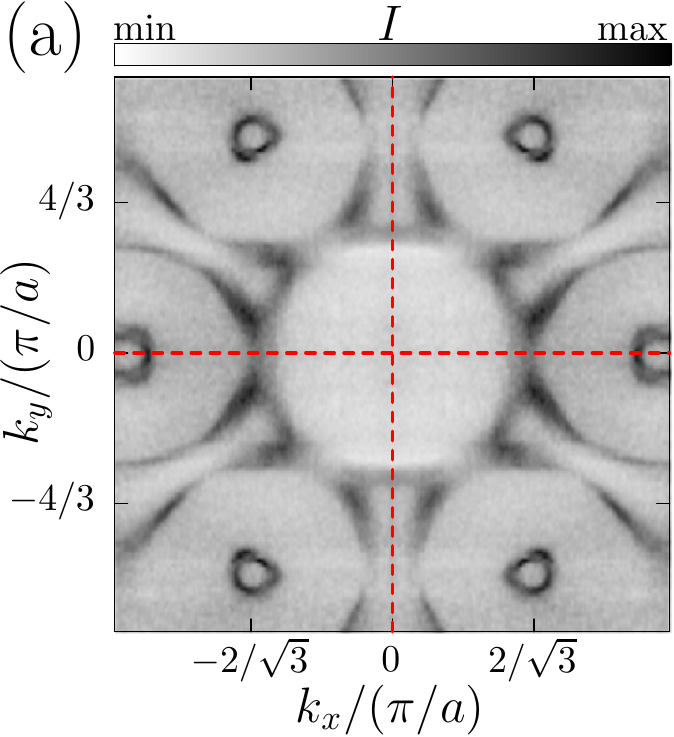}\label{fig:KxKy:a}}\hspace*{.02\columnwidth}
\subfloat{\includegraphics[width=.485\columnwidth]{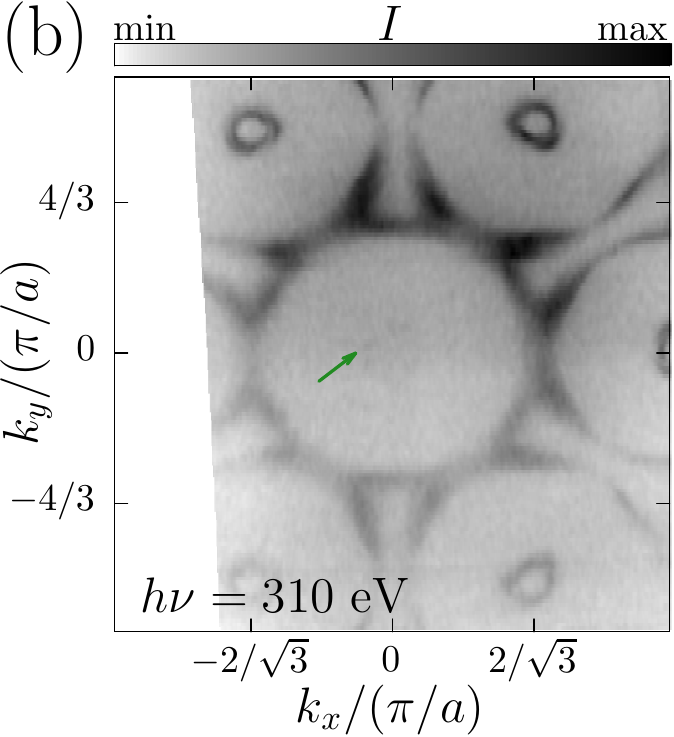}\label{fig:KxKy:b}}\\
\subfloat{\includegraphics[width=.485\columnwidth]{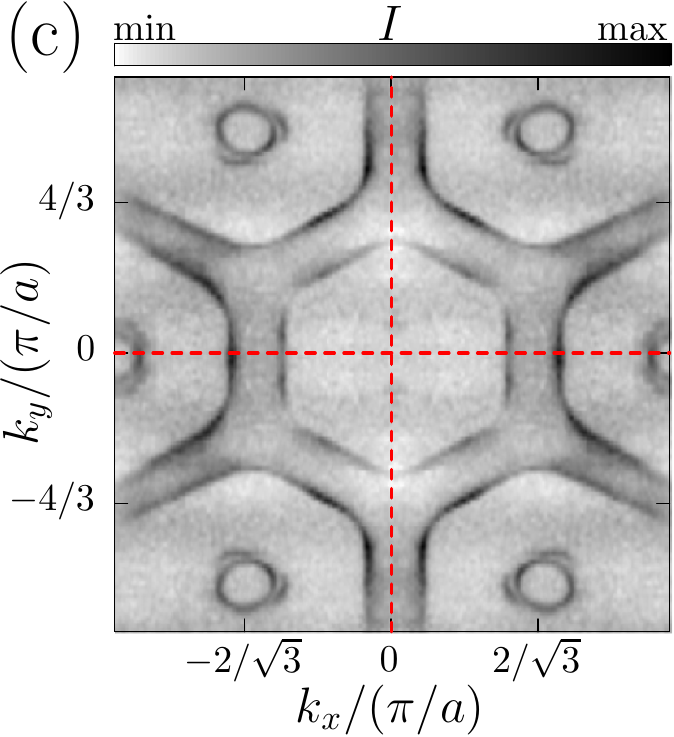}\label{fig:KxKy:c}}\hspace*{.02\columnwidth}
\subfloat{\includegraphics[width=.485\columnwidth]{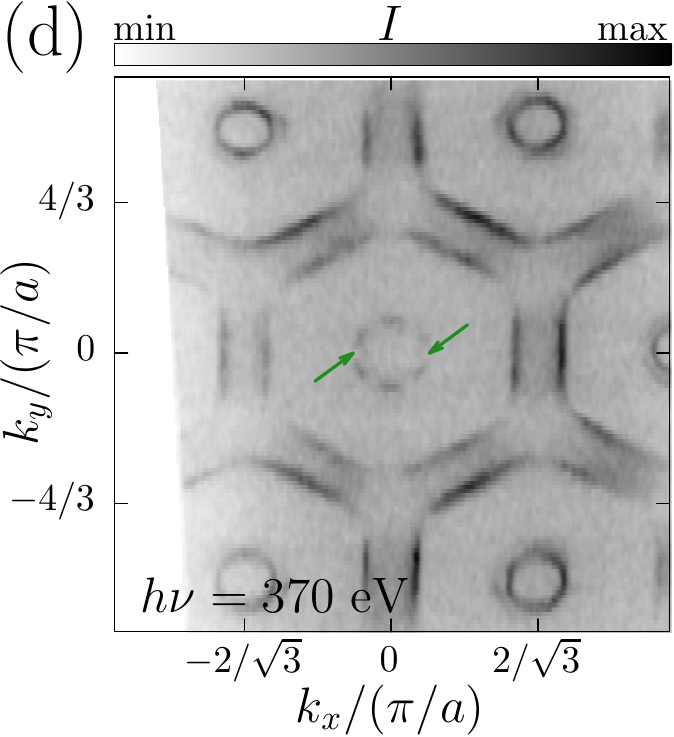}\label{fig:KxKy:d}}
\caption{(a, c) Replotted presentations of Figs.~3(a) and~3(b) without decoration. The red dashed lines border the symmetrized sections of these 
illustrations---only a single quadrant each appears to serve as basis for the presentations. (b,~d)~Corresponding nonsymmetrized data acquired with 
$p$-polarized light with photon energies $h\nu=310$~eV and $h\nu=370$~eV, respectively. The green arrows mark the weakly visible parts of the 
$\sigma$ Fermi-surface cross sections which are absent in Figs.~3(a) and~3(b).}
\label{fig:KxKy}
\end{figure}

Next, taking a closer look at Fig.~2(b) reveals several oddities. The presentation of Ref.~\onlinecite{Sassa-PhysRevB-2015} is redrawn in 
Fig.~\ref{fig:KxKz_GM:a}. In addition to the apparent $k_z$ symmetrization it is noticed that the $\pi$ and $\sigma$ Fermi surfaces feature unnatural 
distinct $k_z$ periodicities. In Fig.~\ref{fig:KxKz_GM:b} only the $\sigma$ part of the shown Fermi surface is replotted using a different grayscale. 
In this view, it becomes clear that the outer $\sigma$ Fermi surface seems to be additionally symmetrized in $k_x$ direction, while the inner one is 
\emph{not}. Moreover, several discontinuities in the data at $k_x = \text{const.}$ seem to be present in Fig.~2(b) [purple arrows in 
Fig.~\ref{fig:KxKz_GM:b}]. These observations are somewhat suspicious and call for a comparison with unsymmetrized data. Unfortunately, according to 
the logbooks of the experiments~\cite{KTH}, \emph{no} measurement corresponding to the situation described in Ref.~\onlinecite{Sassa-PhysRevB-2015} 
(photon-energy scan with \emph{circularly} polarized light covering the $\Gamma$-$M$-$L$-$A$ plane) seems to have been made. However, the authors of 
Ref.~\onlinecite{Sassa-PhysRevB-2015} have conducted corresponding ARPES experiments in different angular ranges with \emph{linearly} ($p$ and $s$) 
polarized light. The resulting Fermi surfaces are depicted in Figs.~\ref{fig:KxKz_GM:c} and~\ref{fig:KxKz_GM:d}. The $\pi$ Fermi-surface cross 
section in the $k_z$ range $14~\pi/c$ to $15~\pi/c$ measured with $p$-polarized light [dotted rectangle in Fig.~\ref{fig:KxKz_GM:c}] bears a striking 
resemblance with the $\pi$ Fermi surface of Fig.~2(b). The small size differences of the electron and hole pockets between Figs.~\ref{fig:KxKz_GM:a} 
and~\ref{fig:KxKz_GM:c} might very well originate from a slightly differently determined Fermi energy~\cite{ETH}. Furthermore, using $p$-polarized 
light, only 
the inner warped $\sigma$ Fermi cylinder is visible in the photoelectron spectrum; only the outer one appears when employing $s$-polarized light. 
Altogether, there are strong indications that Fig.~2(b) contains a somewhat arbitrary combination of different data sets acquired with linearly 
polarized light. This is at substantial variance with the caption of Fig.~2(b) which suggests that a genuine single data set acquired with circularly 
polarized light is shown in Ref.~\onlinecite{Sassa-PhysRevB-2015}.

Finally, the experimental Fermi surfaces shown in Fig.~3 are reviewed. Figures~\ref{fig:KxKy:a} and~\ref{fig:KxKy:c} depict the redrawn presentations 
of Figs.~3(a) and~3(b), respectively. The data here appear to be symmetrized from a single quadrant in the $k_x$-$k_y$ plane using mirror operations. 
The corresponding nonsymmetrized data of single-photon-energy measurements at $h\nu=310$~eV and $h\nu=370$~eV, respectively, are depicted in 
Figs.~\ref{fig:KxKy:b} and~\ref{fig:KxKy:d}. This comparison unearthes yet another peculiarity. The cross sections of the warped $\sigma$ Fermi 
cylinders in the first Brillouin zone (BZ) are essentially nonexistent in the presentations of Ref.~\onlinecite{Sassa-PhysRevB-2015}. The data in 
Figs.~\ref{fig:KxKy:b} and~\ref{fig:KxKy:d}, however, show distinct signs of these parts of the Fermi surface. While it is certainly true that the 
$\sigma$ Fermi surfaces are ``more visible in the second BZ''~\cite{Sassa-PhysRevB-2015}, their complete disappearance in the first BZ particularly 
in Fig.~3(b) [Fig.~\ref{fig:KxKy:c}] remains somewhat questionable. Further examples of equivalent nonsymmetrized experimental Fermi-surface cross 
sections measured using a different geometry, different photon energies, as well as a different light polarization are depicted in Figs.~1.2(a) 
and~1.2(b) of Ref.~\onlinecite{ETH-2011} (with disputable crystal-momentum labels).

\end{document}